\documentclass[aps,prx,10,twocolumn,showpacs,amsmath,amssymb,superscriptaddress]{revtex4-2}

\usepackage{graphicx}
\usepackage{tabularx}
\usepackage{color}

\begin{document}
\title{The remarkable prospect for quantum-dot-coupled tin qubits in silicon}
\author{Wayne M. \surname{Witzel}}
\affiliation{Center for Computing Research, Sandia National Laboratories, 
Albuquerque, New Mexico 87185 USA}
\author{Jesse J. \surname{Lutz}}
\affiliation{Center for Computing Research, Sandia National Laboratories, 
Albuquerque, New Mexico 87185 USA}
\author{Dwight R. \surname{Luhman}}
\affiliation{Sandia National Laboratories, Albuquerque, New Mexico 87185 USA}

\begin{abstract}
Spin-$\frac{1}{2}$ $^{119}$Sn nuclei in a silicon semiconductor could make excellent qubits.  Nuclear spins in silicon are known to have long coherence times.  Tin is isoelectronic with silicon, so we expect electrons can easily shuttle from one Sn atom to another to propagate quantum information via a hyperfine interaction that we predict, from all-electron linearized augmented plane wave density functional theory calculations, to be roughly ten times larger than intrinsic $^{29}$Si.  A hyperfine-induced electro-nuclear controlled-phase (e-n-CPhase) gate operation, generated (up to local rotations) by merely holding an electron at a sweet-spot of maximum hyperfine strength for a specific duration of time, is predicted to be exceptionally resilient to charge/voltage noise.  Diabatic spin flips are suppressed with a modest magnetic field ($>15~$mT for $<10^{-6}$ flip probabilities) and nuclear spin bath noise may be avoided via isotopic enrichment or mitigated using dynamical decoupling or through monitoring and compensation.  Combined with magnetic resonance control, this operation enables universal quantum computation.

\end{abstract}

\maketitle

\section{Introduction} 
\label{sec:intro}
As a potential platform for quantum information applications, solid-state nuclear spins have many desirable properties, including exceptionally-long relaxation times \cite{Steger2012}, fast and precise control \cite{Vandersypen2005}, and promising scalability \cite{Vandersypen2017}. 
Impressive progress has been demonstrated toward fabricating and optimizing nuclear spin systems, among which the most prominent are donors in silicon \cite{Morton2008,Pla2013,Saeedi2013} 
and color centers within diamond \cite{Balasubramanian2009,Maurer2012,Taminiau2014,Metsch2019} or silicon carbide \cite{Bourassa2020}. 

Silicon was recognized as a host material for quantum computing using donor spin qubits by Kane decades ago \cite{Kane1998}, wherein it was argued that a silicon-based platform would eventually outpace competitors by leveraging the myriad fabrication techniques developed in classical microelectronics.
Homogeneous platforms, in which the quantum, classical, and interfacial components all co-inhabit the same host material, have extraordinary engineering advantages at the classical-quantum interface \cite{Reilly2015}, and proposals have recently emerged describing how to operate a scalable two-dimensional qubit system using a transistor-based control circuit and charge-storage electrodes \cite{Veldhorst2017}.
Another crucial boon for silicon is its potential to provide a magnetically-quiet environment.
Naturally-occurring silicon possesses a sparsity ($<5\%$) of finite-spin isotopes, and, by leveraging modern enrichment techniques, one may achieve very low intrinsic nuclear-spin concentrations.
In isotopically-enriched silicon, after eliminating dephasing effects through a schedule of pulse sequences, an electron bound to a hydrogen-like phosphorus donor can maintain coherence on the order of several seconds~\cite{Muhonen2014}.

Nuclear spins controlled with great precision using nuclear magnetic resonance (NMR), together with the long coherence times characteristic of well-separated nuclear spins in enriched silicon, can lead to excellent single qubit performance \cite{Morello2020}.
For quantum computation, however, we must generate entanglement between qubits as well.
In the original Kane architecture this is accomplished through a tunable exchange interaction between bound electrons on neighboring donors~\cite{Kane1998}.
The strong donor confinement potential limits the extent of the bound electron---for example, the prototypical P donor is characterized by an effective Bohr radius of only $1.8$ nm \cite{Smith2017}---introducing challenging fabrication requirements.  This is further complicated by valley-orbit induced exchange oscillations that arise in silicon~\cite{Caldern2008, Gamble2015}.

A promising alternate technique for two-qubit entanglement between arbitrary nuclear-spin qubit pairs involves electron shuttling \cite{Witzel2015,Mills2019,Seidler2021}. 
In this concept, an ancilla electron is initially entangled with one nuclear spin simply through the hyperfine interaction (HFI), in an operation we call electro-nuclear CPhase (e-n-CPhase), and is then coherently transported via an array of quantum dots to interact with a second nuclear spin to achieve long-range nuclear-nuclear entanglement. 
An initial demonstration of coherent spin qubit transport in silicon, critical to this approach, was first demonstrated in 2021 by Yoneda et al. with a promising coherence transfer fidelity of $99.4$\% \cite{Yoneda2021}. 

Donor nuclear spins in silicon represent some of the most coherent qubits available and exhibit a substantial HFI due to the electrostatic confinement \cite{Morello2020}. 
However, the shuttling approach to two-qubit nuclear spin entanglement introduces an additional constraint: the electron must be moved on and off the nuclear spin adiabatically to prevent coherence loss, which may be challenging for a strongly bound electron on a donor.

In contrast, isoelectronic group-IV nuclides pose no obstacle regarding electron shuttling since they do not provide intrinsic electrostatic confinement.
Instead, the confinement of the electron is controlled by electrodes that define a quantum dot.  If an isoelectronic atom resides within the quantum dot, there will be a HFI between an occupying electron and the atom, although weaker than the donor case.
In the case of a $^{29}$Si atom (nuclear spin-1/2), Hensen et al. \cite{Hensen2019} demonstrated that HFI can be prominent enough to initialize, read out, and control single nuclear spins, paving the way for consideration of other isoelectronic species.

While there are a number of naturally-abundant Group IV nuclides with nonzero spin [Table \ref{tab:props}], we find Sn isotopes especially interesting because they are expected to have a strong HFI compared to $^{29}$Si (as shown in Sec.~\ref{sec:DFT}), they are spin-$1/2$, and they are soluble in silicon~\footnote{the maximum solubility of Sn in silicon is approximately $0.016$, similar in magnitude to the Sb donor~\cite{Trumbore1960} and adequate for nuclear spin qubits.}.
Other Group IV nuclides fall short in at least one of these areas. To be specific, $^{13}$C is predicted to have a relatively small HFI (see the appendix)~\footnote{In addition, carbon is among the most common impurities in silicon \cite{Newman1982}, and it has a tendancy to form SiC$_x$ complexes~\cite{Gentile2020} and interstitials~\cite{Platonenko2021}, or, in high concentrations, it precipitates cubic silicon carbide~\cite{Zirkelbach2011}.},
$^{73}$Ge has $I=9/2$ nuclear spin, and $^{207}$Pb has negligible solubility in silicon~\cite{Trumbore1960}.
We note that the $I=9/2$ spin of $^{73}$Ge opens intriguing avenues in quantum information science~\cite{Wang2020,Gross2021} and there has been encouraging progress in the nuclear spin control of $I>1/2$ donors (see, e.g., Refs.~\cite{Asaad2020} or \cite{Morley2010,George2010,Morley2012,Wolfowicz2013,Ranjan2021}, which describe $^{123}$Sb and $^{209}$Bi donor qubits, respectively).
However, the simplicity is attractive for spin-$1/2$ systems as they have no possibility of leakage,
no quadrupole interaction contributing to relaxation \cite{Bloembergen1954}, 
and lend themselves well to electron shuttling and our elegantly simple e-n-CPhase operation.

In this paper we consider the prospect of using Sn incorporated into silicon as a nuclear spin qubit where qubit interactions are achieved through electron shuttling. 
We focus our analysis on two main objectives, with emphasis on the $^{119}$Sn isotope since it has the largest gyromagnetic ratio and greatest natural abundance of the nonzero spin isotopes (Table \ref{tab:props}), although $^{117}$Sn is comparable.
First, we present DFT calculations in Sect.\ \ref{sec:DFT}, which predict that the HFI with $^{119}$Sn will be roughly ten times larger than the HFI with $^{29}$Si. 
This implies a gate time for the entangling e-n-CPhase operation between a Sn nuclear spin and a quantum dot electron of a few microseconds for reasonable quantum dot sizes.
Second, in Sect.\ \ref{sec:2Q_err}, we analyze the probabilities for the important error channels of the e-n-CPhase operation as a function of environmental conditions (quantum dot sizes, magnetic field,
enrichment, and charge/voltage noise in particular).

Our simulations in Sect.\ \ref{sec:2Q_err} suggest that spin-flip errors are greatly suppressed ($<10^{-6}$) with a modest B-field ($>15~$mT), and we use a unique analysis to infer an upper bound on the HFI variability due to charge/voltage noise based upon a comparison between $T_2$ and $T_2^*$ which suggests that a phase-flip error on the nuclear spin qubit during the operation could be below $10^{-7}$ with sufficient control over the quantum dot location owing to a first-order insensitivity to the noise. A phase-flip error on the electron spin is much more significant without extremely high enrichment, but this error can be mitigated using dynamical decoupling or through monitoring and compensation.

The e-n-CPhase gate is a straightforward entangling operation with the potential to be extremely robust (insensitive to noise).
The prospect of high fidelity two-qubit gates together with the prospect of shuttling and the simplicity and proven performance of NMR-driven single qubit operations is remarkably promising for quantum information processing.

\begin{table}[]
    \centering
    \caption{Properties of stable, finite-spin, group-IV nuclides.}

    \begin{tabularx}{1\columnwidth}{cccccc}
    \hline\hline
        Nuclide  & Nuclear & Isotopic & Gyromag.   &Atomic  & Solubility in \\  
                 & spin,   &abund.\footnotemark[1] & ratio,\footnotemark[2]  & radius,\footnotemark[3] & silicon,\footnotemark[4] \\ 
                 & $I$     & (\%)     & $|\gamma_X$/$\gamma_{\rm Si}|$             & $r(X)$/$r$(Si) & $k^{\circ}$=$\chi_{\rm S}$/$\chi_{\rm L}$ \\ 
\hline 
       $^{13}$C    &   1/2   &  1.07 &  1.26  & 0.64 & 5.7 \\     
       $^{29}$Si   &   1/2   &  4.69 &  1.00  & 1.00 & 1.0 \\     
       $^{73}$Ge   &   9/2   &  7.75 &  1.49  & 1.14 & 0.33 \\ 
       $^{115}$Sn  &   1/2   &  0.34 &  1.65  & 1.32 & 0.016 \\    
       $^{117}$Sn  &   1/2   &  7.68 &  1.81  & 1.32 & 0.016 \\
       $^{119}$Sn  &   1/2   &  8.59 &  1.89  & 1.32 & 0.016 \\ 
       $^{207}$Pb  &   1/2   & 22.1  &  1.07  & 1.64 & -- \\ 
       \hline \hline
    \end{tabularx}
    \label{tab:props}
\footnotetext[1]{Isotopic abundances were taken from Ref.\ \cite{Berglund2011}.}
\footnotetext[2]{Gyromagnetic ratios reported relative to $^{29}$Si \cite{CODATA2018}.}
\footnotetext[3]{Atomic radii reported relative to $^{29}$Si \cite{Slater1964}.}
\footnotetext[4]{Solubilities, taken from Ref.\ \cite{Trumbore1960}, are reported in terms of a melting-point distribution coefficient $k^{\circ}$, defined as the ratio 
of the atom fractions of the impurity element in the solid ($\chi_S$) and liquid ($\chi_L$) alloys, respectively. A hyphen designates negligible solubility.}
\end{table}

\section{Hyperfine interaction strengths}
\label{sec:DFT}

As mediator of the primary mode of initializing, addressing, and measuring individual nuclear spins in our scheme,
the HFI is a key factor dictating the feasibility of both single- and inter-site 
nuclear-spin operations. Importantly, Hensen et al.~\cite{Hensen2019} have confirmed 
experimentally that intrinsic $^{29}$Si has a sufficiently strong HFI to facilitate 
a shuttle-based electron nuclear spin approach. Extrinsic defects in silicon, 
on the other hand, have the potential for a stronger HFI, which will, in turn, 
reduce gate times and suppress external noise (e.g., from extraneous nuclear spins). 
In this section we provide an estimate of the HFI for Si:Sn, 
filling an apparent gap in the literature. To this end,
atomic-scale DFT calculations are performed to facilitate comparisons between an
intrinsic $^{29}$Si nucleus and spin-active Sn nuclides residing in a Si host.

\subsection{Theory and methodology} 
\label{sec:DFT_methods}

A HFI occurs when an unpaired electronic spin encounters any nucleus 
possessing a non-vanishing magnetic moment. In such cases
the electronic Hamiltonian is separable as $\mathcal{H} = \mathcal{H}_0 
+ \mathcal{H}_{\rm hf}$, with $\mathcal{H}_0$ describing the field-free 
electronic Bloch states and with the hyperfine Hamiltonian given by
\begin{equation}
    \mathcal{H}_{\rm hf} =  {\bf I} \cdot {\bf A} \cdot {\bf S},
    \label{eq:HFI}
\end{equation}
where {\bf I} and {\bf S} are the nuclear and electronic spin operators
and where {\bf A} is a tensor of coupling terms. 
Truncating at first order in a non-relativistic perturbation expansion \cite{Abragam1989}, 
Eq.\ \ref{eq:HFI} becomes (in SI units)
\begin{eqnarray}
    \mathcal{H}_{\rm hf}^{(1)} = &\frac{2\mu_0}{3} \gamma_e \mu_e \gamma_I \mu_I [{\bf I} 
    \cdot {\bf S}\ \delta({\bf R_{\rm I}})] \nonumber \\
    + &\frac{1}{4\pi r^3} \mu_0 \gamma_e \mu_e \gamma_I \mu_I [\frac{ 
    3({\bf I}\cdot \hat{\bf r})({\bf I} \cdot \hat{\bf r})}{r^2}
    -{\bf I}\cdot {\bf S}],
    \label{eq:general}
\end{eqnarray}
where
$\mu_0$ is the permeability of vacuum, $\gamma_e$ is the electron g-factor, 
$\mu_e$ is the Bohr magneton, $\gamma_I$ is the gyromagnetic ratio,
$\mu_I$ is the nuclear magneton, ${\bf R}_{\rm I}$ 
is the position of the nuclear center, {\bf r} is the 
electron-nucleus distance, and ${\bf I} \cdot {\bf S}\ \delta({\bf R_{\rm I}})$
is the strength of the electron-nuclear spin-spin coupling for nucleus $I$. 

Under conditions of interest, where the electron is moved through electrostatic controls to maximize the HFI, the anisotropic terms are weak 
and {\bf A} is dominated by the isotropic Fermi contact interaction (FCI), which can be obtained 
by integrating the first term of Eq.\ \ref{eq:general} over the electronic
wave function to obtain (in a.u.) \footnote{Eq. \ref{eq:FCI} can actually be derived 
in several ways. Kutzelnigg \cite{Kutzelnigg1988} showed that the $\delta$ function 
appearing in Eq.\ \ref{eq:general} may be avoided entirely, while still remaining 
within a non-relativistic picture, by working within the L{\'e}vy-Leblond theory 
\cite{LevyLeblond1967}. There the FCI is shown to arise naturally in 
the non-relativistic limit of the Dirac equation.}
\begin{equation}
    {\bf A} \approx A_{\rm FCI} =  - \frac{8\pi}{3} \gamma_e \mu_e \gamma_I \mu_I |\Psi({\bf R}_{\rm I})|^2,
    \label{eq:FCI}
\end{equation}
with $|\Psi({\bf R}_{\rm I})|^2$ the electron density at the nucleus.

The bunching factor, a quantity closely related to the FCI, 
was defined by Shulman and Wyluda \cite{Shulman1956} as 
$\eta = |\Psi({\bf R}_I)|^2 / \langle \Psi^2 \rangle_{\rm Av}$, with the denominator 
being the average density taken over the unit cell. It quantifies 
the electron-density enhancement or ``bunching'' at a given nuclear center. 
Van de Walle et al.\ \cite{VandeWalle1993} formulated $\eta$ using spin densities 
($\rho_{\rm spin} = \rho_{\uparrow} - \rho_{\downarrow}$), while Assali et al.\ 
provided a DFT-based procedure for computing $\eta$ for the intrinsic 
$^{29}$Si nucleus in a silicon quantum dot \cite{Assali2011}. 
Their approach generated DFT spin densities on a pristine silicon system 
augmented with an additional conduction-band electron constrained to the
conduction-band edge.
The bunching factor was then calculated as
\begin{equation}
   \eta = \frac{\rho_{\rm spin}({\bf R}_{\rm I})}{\left[\rho_{\rm spin}\right]_{\rm Av}}, 
   \label{eq:eta}
\end{equation}
where $\left[\ldots\right]_{\rm Av}$ is the average spin density in the cell.
In this work we extend the procedure of Assali et al.\ to extrinsic defects in silicon. 

Calculations reported here were performed using full-potential Kohn-Sham DFT 
within a basis of linearized augmented plane waves (LAPW) \cite{Singh1994} 
plus local orbitals \cite{Singh1991},
as implemented in the WIEN2k V19.1 electronic structure software package \cite{Blaha1990,Blaha2020}.
The PBE generalized-gradient approximation was employed to compute the exchange-correlation 
potential within an all-electron formalism of spin-polarized valence and core states.
Scalar relativistic effects were also included, with spin-orbit coupling 
introduced via a separate variational optimization step \cite{MacDonald1980} 
including $p_{1/2}$ basis functions. Core states are treated 
fully relativistically \cite{Desclaux1975}.

Bunching factors reported in the present study were obtained in a three-step process. 
Prepending to the two-step procedure pioneered by Assali et al.~\cite{Assali2011},
a structural optimization was performed first, in which the charge density and nuclear positions
were simultaneously optimized in each self-consistent field (SCF) cycle \cite{Marks2008,Marks2013} 
while excluding spin polarization and spin-orbit coupling. This underlying structure 
was then used to converge spin-polarized, spin-orbit SCF cycles
for the neutral system. Next, an additional
electron was added to the system, accompanied by a uniform positive jellium background
which serves to eliminate inter-image long-range multipole interactions, and, finally,
the Kohn-Sham potential was obtained. As described by Assali et al.,
this procedure effectively constrains the extra electron to a fixed k-point
corresponding to the conduction-band edge of the neutral system.
The desired quantity, $\rho_{\rm spin}$, is computed by summing over 
individual occupied atomic-like alpha and beta spin orbital densities as
$\sum_i |\psi_i^{\alpha} ({\bf r})|^2$ and $\sum_i |\psi_i^{\beta} ({\bf r})|^2$. 
To avoid the nuclear singularity, the contact interaction is estimated
by averaging about a diameter given by the Thomson radius, 
$r_{\rm T} = Ze^2/mc^2$, defined as the distance at which 
the Coulomb energy due to the nuclear charge $Ze$ is equal 
to the electron rest energy in terms of its mass $m$ and the
speed of light $c$. 

\subsection{Results}
First-principles computational modeling is an essential tool for the prediction and 
interpretation of spin-related defect properties observed in silicon \cite{Seo2017} 
and other candidate point-defect qubit materials \cite{Ivdy2018}. For the
intrinsic spin-$\frac{1}{2}\ ^{29}$Si nucleus, simulation has played an 
important role in validating experiments though controversy persists. In 1956, $\eta_{\rm Si}=186\pm18$ 
was obtained from NMR data by Shulman and Wyluda \cite{Shulman1956}, while 
in 1964 the same data were reinterpreted by Wilson resulting in a revised 
value of $\eta_{\rm Si}=178\pm31$ \cite{Wilson1964}. Meanwhile, values of 
$\eta_{\rm Si} \gtrsim 300$ and $\eta_{\rm Si} = 100 \pm 10$ were obtained 
from a 1992 Overhauser-shift \cite{Dyakonov1992} and a 1964 Knight-shift 
measurement \cite{Sundfors1964}, respectively. All-electron DFT calculations
reported by Assali et al.\ predicted the value as $\eta_{\rm Si}=159.4\pm 4.5$
\cite{Assali2011}, which lends credence to the interpretation of Wilson. 
Meanwhile, Philippopoulos et al. implemented a $k\cdot p$ correction on top of 
their DFT calculations and obtained $\eta_{\rm Si} = 88$ \cite{Philippopoulos2020}, 
which agrees better with the 1964 Knight-shift measurement. As the DFT approach 
developed here is inspired by the work of Assali et al., we expect it, too, will 
exhibit good agreement with Wilson's value. However, due to this being only one 
among several measured $\eta_{\rm Si}$ values, it does not convincingly demonstrate 
the accuracy of our approach.

\begin{figure*}
    \centering
    \begin{tabular}{lcr}
    \includegraphics[width=.32\textwidth]{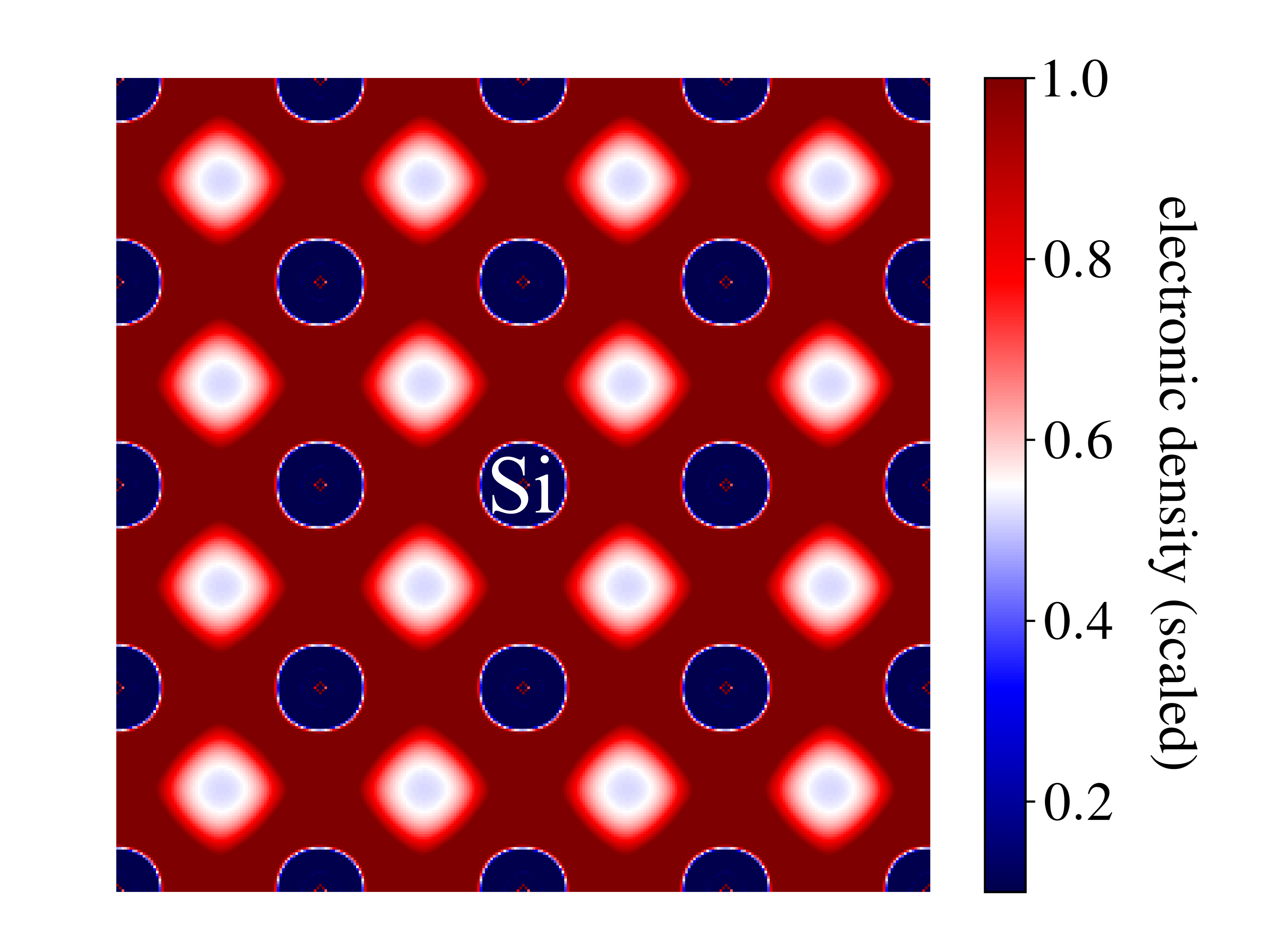} &
    \includegraphics[width=.32\textwidth]{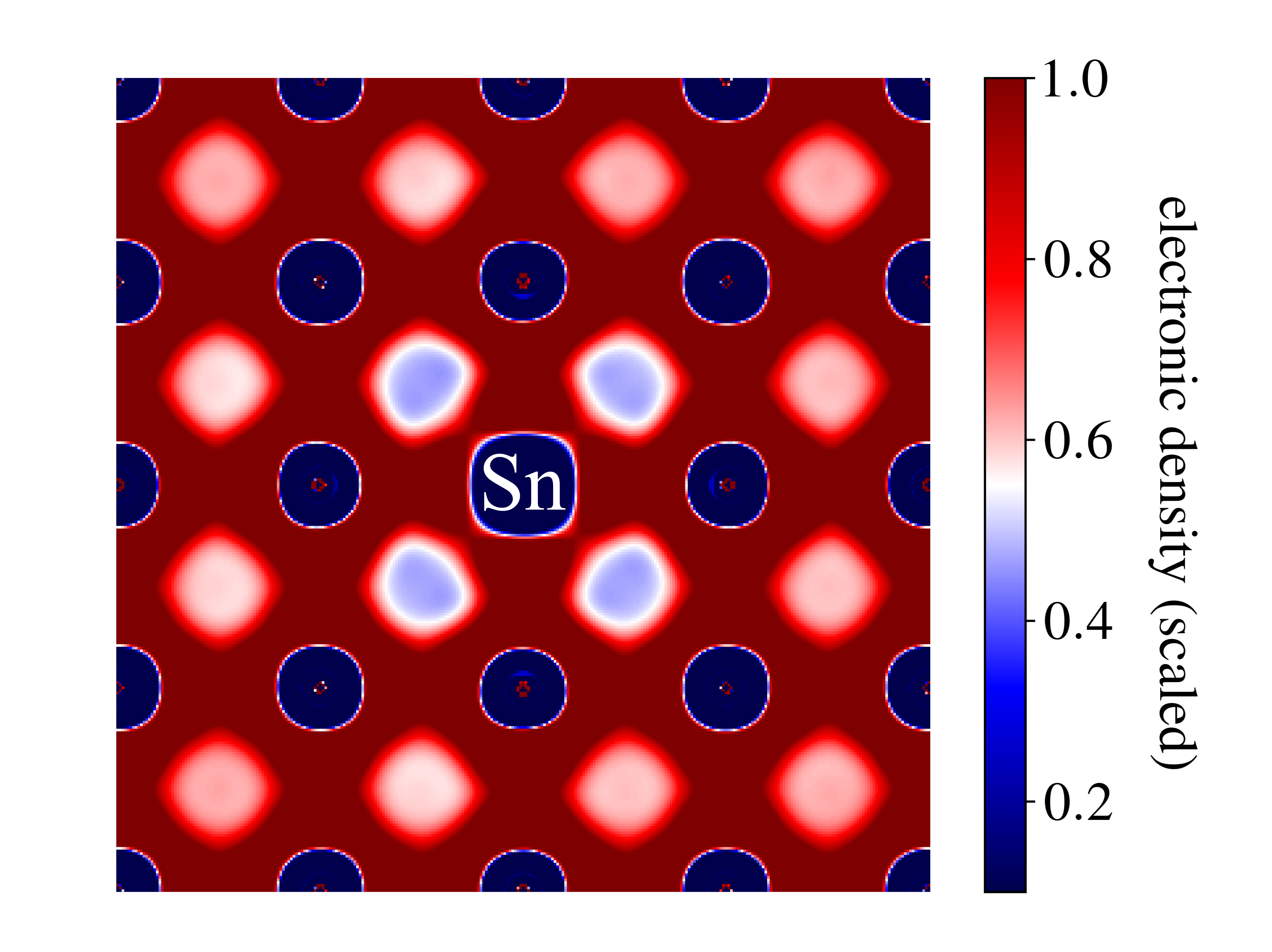} &
    \includegraphics[width=.32\textwidth]{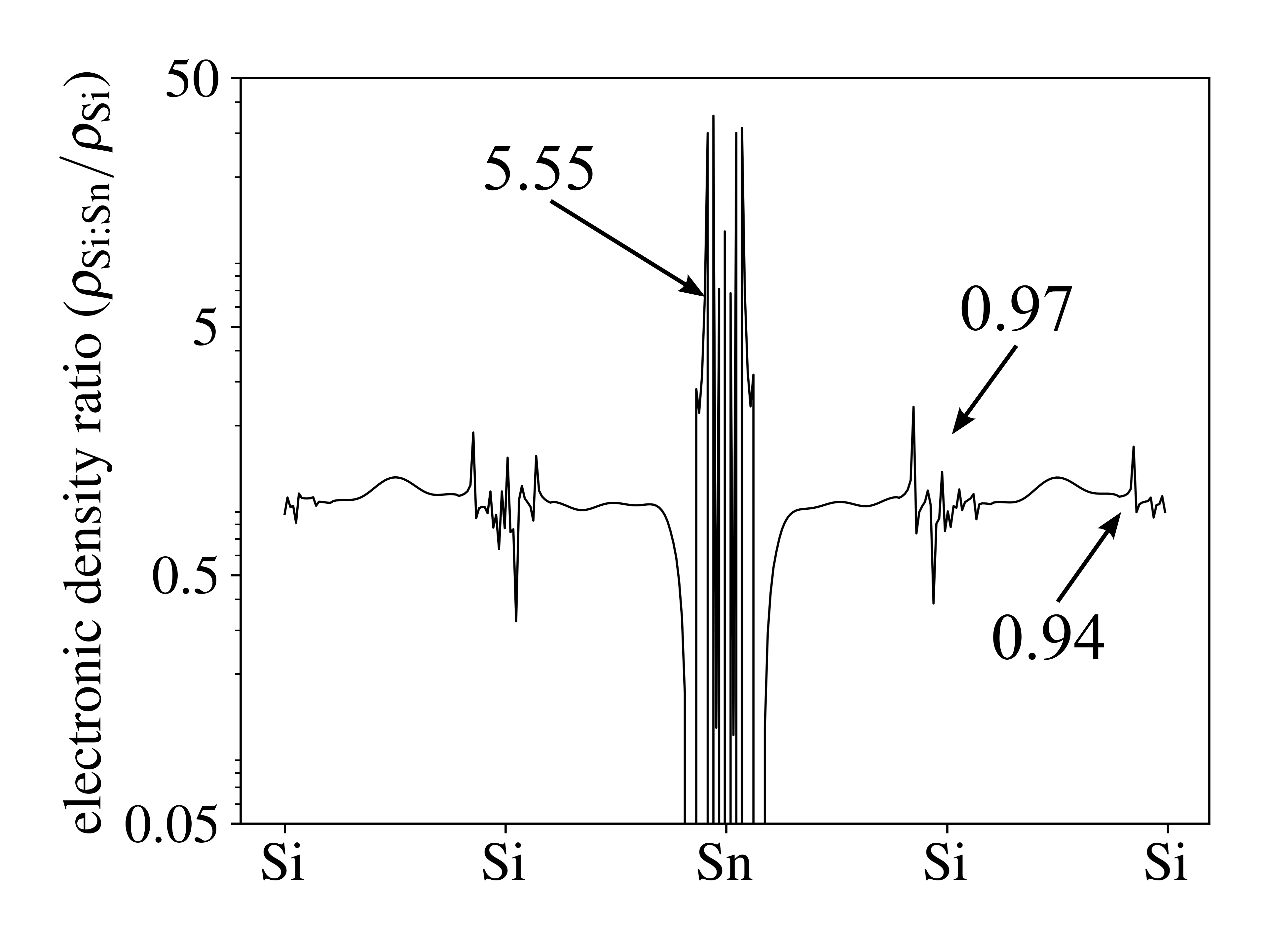}
    \end{tabular}
    \caption{PBE-DFT electron spin densities computed along a $\langle 100 \rangle$ 
    plane of a 4$\times$4$\times$4 supercell for the immaculate Si case (left) 
    and the Si:Sn case (center). In both cases an electron has been introduced and
    constrained to the conduction-band edge. The ratio of a diagonal cross-section 
    of the two densities is also shown (right), where specific values at 
    symmetry-inequivalent nuclear centers are marked for reference. The oscillations
    are attributable to the failure of the plane-wave basis to satisfy Kato's cusp condition.
    }
    \label{fig:densities}
\end{figure*}

Due to the controversy surrounding the accepted value for $\eta_{\rm Si}$, 
here we seize upon the opportunity to further validate against an unambiguous 
reference value provided by Kerckhoff et al.\ \cite{Kerckhoff2021} in 2021. 
There a value of $\eta_{\rm Ge}=570\pm171$ was obtained by leveraging noise 
spectra measured for the Si:$^{73}$Ge system. The 30\% uncertainty associated 
with this value, which seems large at first, is quite comparable to the spread 
of experimental $\eta_{\rm Si}$ values. Therefore, in preparation for computing the target value for Sn, $\eta_{\rm Sn}$, we first computed $\eta_{\rm Si}$ and $\eta_{\rm Ge}$.
The reason for this is twofold. In addition to validating the applicability 
of our procedure for Si:Sn, demonstrating agreement for a second benchmark system 
bolsters confidence that our $\eta_{\rm Si}$ is of similar accuracy, thereby 
adding another data point toward the adoption of an accepted value for $\eta_{\rm Si}$.

As a first check of our computational procedure, we sought to confirm that a 
conduction electron will indeed have an increased probability density at the
Sn sites, without being too strongly localized and donor-like.  
That is, we do not want wave-packet localization to disrupt the ability 
to smoothly move a quantum dot electron with electrostatic controls. 
Figure \ref{fig:densities} shows the scaled electronic density of a 
4$\times$4$\times$4 Si supercell for pristine Si, a defective Si:Sn 
system, and the ratio of the two along a diagonal line cut. Note that
the Sn impurity slightly modifies the interstitial region adjacent to Sn
(faint blue), but it does not significantly deform the density of the adjacent Si sites 
(deep blue) relative to the more distant neighbors. This is quantified 
in the ratio plot where we indicate the values at the Sn and Si site locations.
As compared with a Si nucleus in the pristine bulk, the density at the 
substitutional Sn nucleus is over five times higher. Meanwhile the 
neighboring Si contact densities are changed by only a few percent. 
This distinguishes an isoelectronic defect from a donor. It is thus 
expected that introducing a Si:Sn impurity will only
weakly impact the extent of the envelope of the electronic wavefunction.

\begin{figure}
    \centering
    \includegraphics[width=\linewidth]{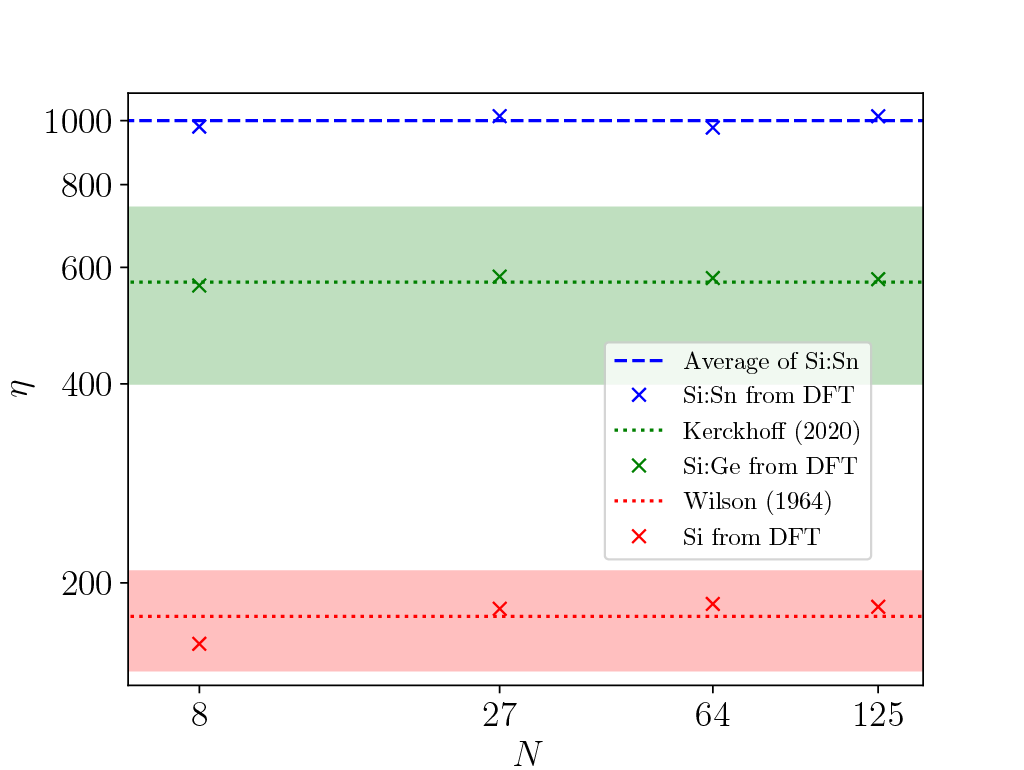}
    \caption{Fermi contact densities obtained by DFT calculations using the PBE functional,
    as performed on various $n\times n \times n$-dimensional supercells with atom count $n^3=N$. 
    Included for reference are the measured values $\eta_{\rm Si}=178\pm 31$
    and $\eta_{\rm Ge}=570\pm 171$, corresponding to a $^{29}$Si nucleus in the bulk \cite{Wilson1964}
    and a substitutional $^{73}$Ge defect in silicon \cite{Kerckhoff2021}, respectively.
    Shaded bands represent the experimental uncertainties.}
    \label{fig:etas}
\end{figure}

Next, we moved to compute $\eta$ values as defined in Eq.\ \ref{eq:eta}, which
unlike the density ratios above, require computation of electron-spin 
densities for a single system only.
We performed convergence studies with respect to both supercell size 
and the number of $k$ points, with computations performed on $n^3$-atom 
supercells having $n=2,3,4,5$ using integration grids containing 
between 8 and 2000 $k$ points. Figure \ref{fig:etas} collects the 
computed $\eta$ values for $^{29}$Si, $^{73}$Ge, and $^{119}$Sn defects 
in silicon, showing for comparison the experimental $\eta_{\rm Si}$
value of Wilson and the $\eta_{\rm Ge}$ value of Kerckhoff et al.\ 
Good agreement was found between the measured and theoretical values
for both $\eta_{Si}$ and $\eta_{\rm Ge}$, which bolsters confidence 
in the accuracy of our computed value of $\eta_{\rm Sn} = 996.4$, 
obtained as an average over several supercell sizes. Alternatively
this may be expressed as a ratio with respect to $^{29}$Si in
immaculate silicon as $\eta_{\rm Sn} = 5.6\eta_{\rm Si}$,
which is virtually identical to the density ratio at the Sn nucleus shown
in Figure \ref{fig:densities}. Subsequent application of the gyromagnetic ratios 
(see Table \ref{tab:props}), returns a HFI enhancement factor of over $10$ 
for Si:$^{119}$Sn, as compared with intrinsic Si:$^{29}$Si.
Furthermore, by including a standard correction for relativity in Eq.\ \ref{eq:FCI},
the Si:Sn absolute HFI grows to 2400, corresponding to an enhancement
of 13.5 times the value of intrinsic $^{29}$Si (see the appendix for details).
As a conservative estimate for the analysis that follows, 
we adopt the non-relativistic enhancement factor of 10.

\begin{figure}
    \centering
    \includegraphics[width=\linewidth]{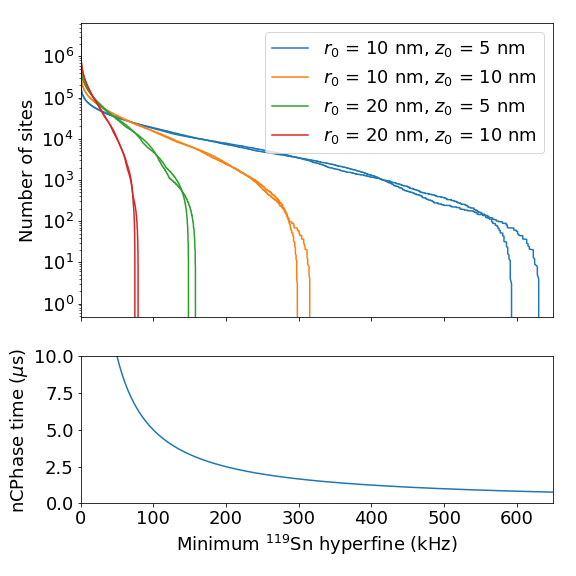}
    \caption{(Top) Number of lattice site locations at which a $^{119}$Sn would have a hyperfine interaction 
    above the corresponding minimum strength in frequency units (energy multiplied 
    by Planck's constant) for various quantum dot shapes and valley phase extremes 
    using our estimated value of $\eta_{\rm Sn} = 996.4$.  We used the simple 
    wavefunction model of Eq.~\ref{eq:wavefunction} as a basic characterization 
    of quantum dot sizes.  The two curves for each color correspond to the extreme valley
    phases of $\theta_v \in \{0, \pi\}$ where $\theta_v = 0$ yields the largest 
    hyperfine strength at the vertical center of the quantum dot. (Bottom) Minimum 
    (i.e., in the limit of instantaneous hyperfine interaction on/off switching) 
    e-n-CPhase gate time in correspondence with each hyperfine strength 
    ($0.5/A h$ where $A$ is the hyperfine energy and $h$ is Planck's constant).
    \label{fig:hf_distribution}}
\end{figure}

To compute actual HFI strengths for individual nuclei, the envelope of the quantum dot wavefunction must be known.  
As a convenient proxy, we use a simple model that derives from a well with infinite barrier 
and parabolic lateral confinement at zero electric field.  At each nuclear site, $n$, with a bunching factor of $\eta_n$, our proxy wavefunction,
parameterized by a radius $r_0$ and thickness $z_0$, is
\begin{eqnarray}
    \nonumber
    |\Psi(n)|^2 &\propto&
        \eta_n e^{-((x_n-x_0)^2 + (y_n-y_0)^2) / r_0^2} 
        \cos^2{\left(\frac{z_n \pi}{z_0}\right)} \\
    & & \times \cos^2{\left(k_0 z_n - \theta_v / 2\right)},
    \label{eq:wavefunction}
\end{eqnarray}
where $\theta_v$ is the valley phase.  Since the valley phase of quantum dot may depend upon 
local chemical details of the quantum dot environment and its interfaces, 
we treat it as an independent parameter in our model.
The valley oscillation frequency is based upon effective mass theory for silicon, 
$k_0 = 0.85 \cdot 2 \pi / a_0$ with $a_0 = 0.543$~nm as a standard lattice constant 
for Si~\cite{CODATA2018}. The form of this last valley-dependent factor is dictated 
by the symmetry of the bulk silicon lattice (with translation + inversion symmetry).
Using this model, our estimate of $\eta_{\rm Sn} = 996.4$, and the gyromagnetic ratio 
of Sn, Fig.~\ref{fig:hf_distribution} shows the distribution of hyperfine interaction strengths 
at all possible sites for a few different quantum dot shapes, 
as well as corresponding minimum gate times for an e-n-CPhase operation.

\section{Two-qubit error channels}
\label{sec:2Q_err}

In this section, we consider the errors incurred during the e-n-CPhase gate operation between an electron spin qubit and Sn nuclear spin qubit.  
For a complete error model, one must also study electron spin preparation and measurement, 
shuttling of individual electrons, as well as ESR and NMR single-qubit rotations, 
all of which have all been demonstrated experimentally with promising results~\cite{Pla2013,Muhonen2014,Kawakami2016,HarveyCollard2017,Yang2019,Mills2019,Seidler2021,Yoneda2021}.
Our theoretical analysis suggests that two-qubit operations between transportable electrons and stationary Sn nuclei 
can have exceptionally good fidelities, holding great promise as a quantum information processing technology.

Up to local $Z$ rotations, the e-n-CPhase gate is straightforward to implement in the presence of a finite magnetic field.  Start with the electron away from the nuclear spin qubit such that their interaction is negligible.  Next, adiabatically move the electron to maximize the HFI with the target nucleus and hold the electron there for a specific duration of time.  Finally, adiabatically move the electron away again.  In the adiabatic limit, the operation must be diagonal in the original eigenbases of the two spins (with the quantization axis predominantly determined by the external magnetic field).
Assuming the transit duration is negligible compared with the holding duration, this operation will induce a controlled-$Z$ rotation component in the original eigenbases that is approximately linear in the holding duration.  Setting this duration for a rotation of $\pi$ will generate the e-n-CPhase operation apart from an inconsequential global phase and local $Z$ rotations (accounting for the four degrees of freedom of the diagonal unitary operation).  We may compensate for systematic local $Z$ rotations through single qubit rotations effected by magnetic resonance pulses.  In this section, we focus on errors incurred during the two-qubit entangling operation described above (independent of single qubit rotations during the magnetic resonance pulses).

The contact HFI is short range (being proportional to the electron density at the nucleus) 
and much stronger than the longer range dipolar interaction (which is below 8 Hz at a 20 nm distance 
and scales inversely with distance cubed).  Therefore, we can regard the interaction between a Sn nucleus and an 
electron to effectively be switched off except during the time when the electron is in close 
proximity of the Sn qubit for an intended operation.  For comparison, the dipolar interaction between electrons 
is about 1800 times stronger (13 kHz at 20 nm, or 13 Hz at 200 nm).  However, if a proper distance 
is maintained between different electron qubits, and the electron qubits are relatively short lived, 
these interactions can be neglected.  Having gyromagnetic ratios $<1000$ times smaller than electrons 
(see Table \ref{tab:props}), the Sn qubits should be relatively well isolated from most other sources 
of magnetic noise, which can also be mitigated using spin echo pulses that can greatly extend nuclear spin qubit lifetimes~\cite{Witzel2010,Pla2013,Muhonen2014,Morello2020}. 

Thus, the main errors of concern involving interactions between qubits occur during the two-qubit operations 
and should be independent if a 
sufficient distance between electrons is maintained.  Furthermore, if the electrons are transient
and do not have too many interactions with nuclear spins (e.g., they are used solely for mediating 
gates between nuclear spins such as the inter-nuclear CPhase gate described in Eq.~(6) of
Ref.~\onlinecite{Witzel2015}), correlations of these errors between different operations 
should not be a major concern.  For this reason, we report error estimates based upon 
Born-rule probabilities since there is little opportunity for coherent errors between 
different operations to add (constructively or destructively). 
If coherent errors do add in a systematic and controllable way, 
it should be possible to exploit this and adjust the schedule in order to cancel the coherent 
errors instead.  We therefore feel justified in reporting error probabilities rather than amplitudes; 
however, much depends upon the details of quantum circuit schedules,
which is beyond our current consideration.

One important error category involves electron orbital and/or valley excitations.
If such an excitation occurs, it can induce an uncertainty in the HFI with the nuclear spin qubit and render the two-qubit operation to be unreliable.  If the electron does not relax quickly, it could induce errors on every Sn that this electron touches.
This is mitigated with sufficient orbital and valley energy gaps that are device specific.  The orbital energy gap is determined by the electrostatic confinement of the quantum dot.  The valley energy gap can be made large in a Si-MOS quantum dot with a strong vertical field to force the electron against the interface~\cite{Yang2013, Gamble2016} and can be made large in Si/SiGe devices with alloy engineering~\cite{wuetz2021atomic, mcjunkin2021sige}.

Maintaining large orbital and valley energy gaps and performing smooth electron shuttling operations is clearly important for good qubit operation fidelities.  In the following discussion, we will consider the remaining errors assuming the electron follows the ground state faithfully.
Specifically, we address the fifteen two-qubit Pauli error channels for these two spin-$1/2$ particles:
\begin{enumerate}
    \item Electron and/or nuclear spin flip.  Assuming that contact HFI dominates
    over any anisotropic interaction, the most likely error of this type would be a correlated 
    flip-flop error  via a diabatic transition from a sudden change of the contact HFI: $\hat{X} \otimes \hat{X}$, $\hat{X} \otimes \hat{Y}$, $\hat{Y} \otimes \hat{X}$, and $\hat{Y} \otimes \hat{X}$.  Including single flip errors that we anticipate to be less likely, this accounts for twelve of the fifteen error channels.
    \item $\hat{Z} \otimes \hat{Z}$ error.  This is caused by an uncertainty in the time integration 
    of the HFI with the Sn qubit.  With relatively slow gate times, uncertainty of the peak hyperfine 
    strength will likely dominate over timing jitter.
    \item Electron $Z$ error.  This is caused by uncertainty in the effective magnetic field experienced 
    by the electron due to sources other than the Sn qubit.  This will likely be dominated by the 
    Overhauser field induced by extraneous nuclear spins.
    \item Nuclear $Z$ rotation.  This will likely be negligible if dynamical decoupling is employed 
    to cancel slowly varying magnetic fields by using, for example, a protocol such as described in Ref.~[\onlinecite{Witzel2015}]).
\end{enumerate}

The following three subsections are dedicated to providing a deeper analysis of items 1, 2, and 3, respectively. 
The last will likely be negligible in comparison with NMR rotation 
errors of the Sn qubits.

\subsection{Diabatic flip-flops}
\label{sec:diabatic}
\begin{figure}
    \centering
    \includegraphics[width=\linewidth]{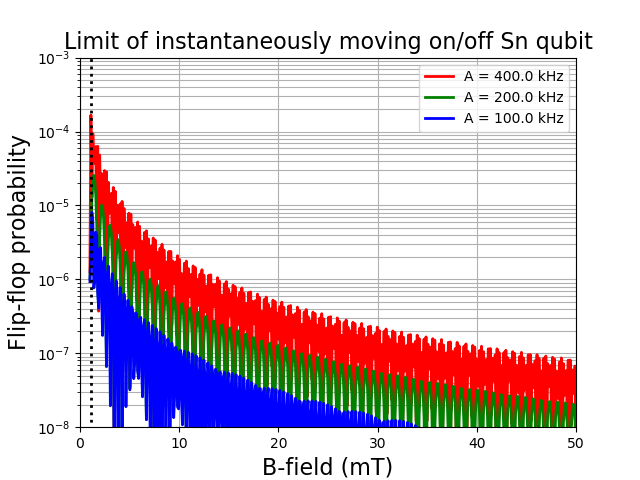} \\
    \includegraphics[width=\linewidth]{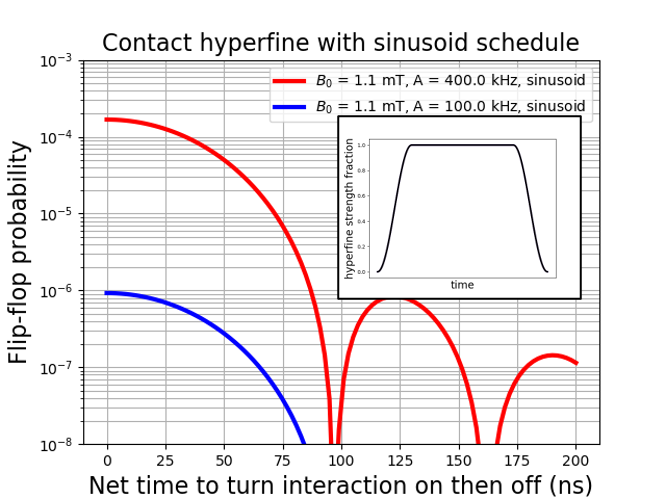}
    \caption{(Top) Diabatic flip-flop probabilities as a function of external B-field 
    for three different HFI strengths taken in the limit of instantaneous
    HFI on/off switching. A vertical dotted line marks $B_0 = 1.1$~mT which is used in the bottom plot.
    (Bottom) Diabatic flip-flop probabilties as a function of net on/off 
    switching times at $B_0 = 1.1$~mT using a sinusoid shape for the hyperfine transitions 
    as depicted in the inset.  Probabilities can improve by using slow and smooth transitions, but this is not necessary if a modest B-field strength is used.}
    \label{fig:flip_flop_probs}
\end{figure}

Given long $T_1$ times of electron spins in silicon~\cite{Tyryshkin2003}, 
the dominant spin-flip error mechanism during an e-n-CPhase operation is expected to be a correlated flip-flop 
($\hat{X} \otimes \hat{X}$) induced via a diabatic transition while switching 
the contact HFI on/off. The HFI is effectively 
turned on/off by moving the electron on/off the Sn qubit via electrostatic controls.  
Using QuTiP~\cite{Johansson2012}, we have simulated various scenarios using a time-dependent Hamiltonian to switch on/off the HFI: $\hat{H}(t) = A(t) \hat{I} \cdot \hat{S}$.
The limit of instantaneous switching is a worst-case scenario.  
However, this worst-case error probability is exceptionally low in the presence of a modest 
external B-field as shown at the top of Fig.~\ref{fig:flip_flop_probs}.  
The error probability is below $10^{-6}$ for all of our HFI 
strengths beyond a modest 15~mT B-field, well below known quantum-error-correction thresholds~\cite{Stephens2014}. 
On the bottom figure, we do see that error probabilities can be reduced even further
(enabling even smaller B-fields at these very low error rates), by using slow and smooth transitions. 
In either case, this error mechanism is not a significant concern owing to the expected
HFI of the Sn qubit being very weak relative to the electron and nuclear Zeeman-energy difference 
at modest (tens of mT) B-fields.

\subsection{Correlated $\hat{Z} \otimes \hat{Z}$}
\label{sec:zz_err}

One of the main advantages of using nuclear spin qubits with interactions 
mediated by electron spins, besides minimal crosstalk concerns and precise NMR/ESR control, 
is that we can, in principle, take advantage of a ``sweet-spot" in the electro-nuclear interaction 
provided we are able to move the electron to maximize the HFI and minimize its uncertainty.
That is, 
by maximizing the HFI, we become insensitive to control uncertainty 
and charge noise to first order in a perturbative expansion. 

Disregarding (i.e., projecting away) the other error channels, and assuming that timing jitter is negligible (given the relatively long expected operation time on the $\mu$s scale), the gate time can be tuned for the e-n-CPhase gate operation to become $\hat{U} = e^{i \phi (\hat{Z} \otimes \hat{Z}) / 2} = \cos{\left(\phi / 2\right)} \hat{I} \otimes \hat{I} + i \sin{\left(\phi / 2\right)} \hat{Z} \otimes \hat{Z}$ where $\phi = \left(1 + \frac{\Delta A}{A}\right) \pi$ with $\Delta A$ representing the uncertainty in the HFI.  The Born-rule probability of a correlated $\hat{Z} \otimes \hat{Z}$ phase flip error after an e-n-CPhase operation is therefore $\cos^2{(\phi/2)} = \sin^2{((\phi-\pi)/2)} \approx 
\left(\frac{\pi}{2}\right)^2 \left ( \frac{\Delta A}{A} \right )^2$.  To lowest order, the average error probability is
\begin{equation}
\label{eq:zerr_via_deltaA}
P_{Z \otimes Z}^{\rm err} \approx \left(\frac{\pi}{2}\right)^2 \left \langle \left ( \frac{\Delta A}{A} \right )^2 \right \rangle
\end{equation} 
where the angle brackets denote averaging over noise realizations that impact $\Delta A$.

In the analysis that follows, we show that the $T_2 / T_2^{*}$ ratio can actually serve as a proxy to determine expectations for $P_{Z \otimes Z}^{\rm err}$, via $\left \langle \left ( \frac{\Delta A}{A} \right )^2 \right \rangle$ and Eq.~\ref{eq:zerr_via_deltaA}, under a few simplifying assumptions.
First, we assume that
\begin{itemize}
    \item We have high-precision control of the quantum dot wavefunction in both lateral directions;
    \item The charge noise of the system does \emph{not} significantly influence the quantum dot in any manner that is fundamentally different from this lateral control.
\end{itemize}
The former requires more than a linear array of electrodes; at least one additional electrode would be required to move the electron in a direction that is orthogonal to a linear quantum dot array.  The second assumption is potentially violated by the fact that a vertical field or local charge fluctuation can perturb the valley phase.  We assume that such effects are negligible, however.  This is not unreasonable.  For Si/SiGe quantum dots, the fixed alloy composition largely dictates the valley phase \cite{Goswami2006,Friesen2007,Friesen2010,Borselli2011,Jiang2012,Neyens2018,wuetz2021atomic, mcjunkin2021sige}.  For Si-MOS quantum dots, the position of the oxide interface largely dictates the valley phase, given a sufficient vertical electric field \cite{Yang2013,Gamble2015,Gamble2016}.

The effect of vertical fields and local fluctuations deserves scrutiny in future work, but we use the simplifying assumptions above for the analysis presented here.
Furthermore, we take Eq.~\ref{eq:wavefunction} as the form of the quantum dot wavefunction, parameterized by $x_0$, $y_0$, and $\theta_v$.  
By our second assumption above, $\theta_v$ only depends upon $x_0$ and $y_0$.
We now consider a perturbation of the $x_0$ and $y_0$ parameters.  Without loss of generality, we take $x_0 = y_0 = 0$ (absorbing them into $x$ and $y$).  For convenience in notation, let $\xi_0 = x_0$ and $\xi_1 = y_0$.  Since the HFI is proportional to the electron probability density at the nuclear site, $|\Psi(n)|^2$, to second order we have
\begin{eqnarray}
\label{eq:A_perturbation}
\frac{\Delta A}{A} &=& \sum_i c_i \Delta \xi_i + \sum_{i, j} c_{i, j} \Delta \xi_i \Delta \xi_j
+ \mathcal{O}\left(\left(\Delta \xi\right)^3\right) \\
\nonumber
c_{0} &=& \left(\frac{2 x}{r_0^2} + \tan{\theta(z)} 
\frac{\partial \theta_v}{\partial x_0} \right) \\
\nonumber
c_{1} &=& \left(\frac{2 y}{r_0^2} + \tan{\theta(z)} 
\frac{\partial \theta_v}{\partial y_0} \right) \\
\nonumber
c_{0,0} &=& \frac{2 x^2}{r_0^4} - \frac{1}{r_0^2} + \frac{\tan{\theta(z)}}{2} 
\left(\frac{\partial^2 \theta_v}{\partial x_0^2} + \frac{4 x}{r_0^2} \frac{\partial \theta_v}{\partial x_0}\right) \\
\nonumber
&& {} + \frac{\tan^2{\theta(z)} - 1}{4} \left(\frac{\partial \theta_v}{\partial x_0} \right)^2
\end{eqnarray}
for $|z| < z_0 / 2$ (otherwise $A$ is zero in our model) where $\theta(z) = k_0 z - \theta_v/2$.  The form of $c_{1,1}$ is similar to $c_{0,0}$, and $c_{0,1}$ and $c_{1,0}$ will not be important if we assume that $\Delta \xi_0$ and $\Delta \xi_1$ express  independent random variables.

If the dot can be moved relative to a target qubit at $(x, y, z)$ such that the first order term of $\frac{\Delta A}{A}$ vanishes (i.e., the sweet spot), then $2 x = -r_0^2 \tan{\theta(z)} \frac{\partial \theta_v}{\partial x_0}$ and $2 y = -r_0^2 \tan{\theta(z)} \frac{\partial \theta_v}{\partial y_0}$ so that
\begin{eqnarray}
\label{eq:perturb_sweetspot}
c_0 &=& c_1 = 0 \\
\nonumber
c_{0,0} &=& -\frac{1}{r_0^2} + \frac{\tan{\theta(z)}}{2} 
\frac{\partial^2 \theta_v}{\partial x_0^2} -
\frac{\tan^2{\theta(z)} + 1}{4} \left(\frac{\partial \theta_v}{\partial x_0} \right)^2 
\end{eqnarray}
Furthermore, we can be selective with our choice of nuclear qubits at the expense of reducing their density (i.e., increasing the average distance between qubits in the chip).  In our analysis, we choose to select only qubits for which $\tan^2{\theta(z)} \leq 1$; assuming $\theta_v$ is distributed evenly in this respect, this selectivity only reduces the candidates by half.  

We will now show how $T_2$-experiments (Hahn~\cite{Hahn1950} or CPMG~\cite{Carr1954, Meiboom1958}) can inform $P_{Z \otimes Z}^{\rm err}$.
More specifically, we will relate $\left \langle T_2^* / T_2 \right \rangle$ to $\left \langle \frac{\Delta A}{A} \right \rangle$ which determines $P_{Z \otimes Z}^{\rm err}$ via Eq.~\ref{eq:zerr_via_deltaA}.
Although spin echo experiments of quantum dots are typically limited by the flip-flop dynamics of the nuclear spin bath~\cite{Witzel2012,Jock2018}, they will also be sensitive to shifts of the wavefunction that alter HFIs, serving as a bounding probe of electron location reproducibility in the presence of charge noise.  That is, long $T_2$ spin echo lifetimes would not be possible without the ability to control the location of electrons enough to keep $\left \langle \left(\Delta A_n / A_n \right)^2 \right \rangle$ small for the background of nuclear spins labeled by $n$.

These experiments may be performed in a single- or double-electron setting.  Using two electrons is sensible since we can use Pauli-spin blockade readout~\cite{Johnson2005} which does not require the large magnetic field needed for single-spin readout~\cite{Elzerman2004}.  Furthermore, refocusing pulses for the spin echo with two electrons may be performed using exchange-based swaps rather than requiring ESR.  In this setting, echo experiments amount to preparing a singlet state (the ground state when two electrons are loaded into a confined space), swapping electron spins during their lifetime to balance the amount of time they each spend in particular locations, and then reading singlet versus triplet via Pauli-spin blockage to determine the remnant of singlet/triplet rotations that were not canceled through swapping (as well as spin flip errors).   

From $T_2$ experiments, we can bound the contributors to $\left \langle \left(\Delta A / A \right)^2 \right \rangle$.  
$T_2^*$ measurements, in the ergodic~\footnote{While it can take a very long lime to reach the ergodic $T_2^*$, particularly in experiments with enriched silicon, this could be accelerated by using NMR to rotate the nuclear spins allowing one to sample a greater variety of states more quickly.} limit, are also useful for obtaining this error probability bound.
While $T_2$ is sensitive to changes of the HFIs (in addition to nuclear flip-flops), $T_2^*$ is sensitive to the magnitudes of the HFIs.  As we will show, the $T_2^*/T_2$ ratio provides a robust proxy to the $\hat{Z} \otimes \hat{Z}$ error probability bound.

Let the $\hat{\phi}$ quantum operator represent the net Overhauser rotation induced during the experiment (reversed with each refocusing pulse).  In the limit of a large number of nuclear spins, the outcomes are Gaussian-distributed by the central limit theorem.  The echo is the difference in averaged measurement outcomes.
For an upper bound of the echo decay curve, we consider the limit in which the nuclear spin polarizations are static and pulses and measurements are instantaneous and ideal.  Thus,
\begin{eqnarray}
\nonumber
{\rm Echo} &\leq& \left \langle \cos^2{(\hat{\phi}/2)} - 
   \sin^2{(\hat{\phi}/2)} \right \rangle  = 2 \left \langle \cos^2{(\hat{\phi}/2)} \right \rangle -1 \\
    &=& \exp{\left(-\left \langle \hat{\phi}^2 \right \rangle /2 \right)} \equiv e^{-\left(t / T_2^{(X)}\right)^2 }, \\
\label{eq_T2_sqrd}
\left(T_2^{(X)}\right)^2 &\leq& 2 t^2 / \left \langle \hat{\phi}^2 \right \rangle.
\end{eqnarray}
The $X$ in $T_2^{(X)}$ is a placeholder to mark the type of experiment (the pulse sequence).

A CPMG pulse sequence with $m$ refocusing pulses is a sequence that, apart from details about the initial and final $\pi/2$ rotations that are unimportant here, may be expressed as $(\tau \rightarrow \pi \rightarrow \tau)^m$.  Each $\tau$ denotes free evolution for time $\tau$, $\pi$ denotes a refocusing pulse, and exponentiation by $m$ denotes repetition.  A Hahn echo, for our purposes, is simply the $m=1$ case of CPMG.  
For simplicity, we assume an independent noise realization of $\Delta A$, via $\Delta x_0$ and $\Delta y_0$, for each free evolution time.  In reality, the HFI may vary during the free evolution time, but we can lump that into an effective uncertainty.  Also, the noise realizations may be correlated as a function of time; for this reason, our analysis only really informs $P_{Z \otimes Z}^{\rm err}$ over the $T_2$ timescale since the last time that the e-n-CPhase gate was tuned up.
With $m \geq 1$, and assuming a decay dominated by spin $1/2$ nuclei (e.g., $^{29}$Si) in addition to independent noise realizations of $\Delta A$,
\begin{eqnarray}
\nonumber
\left \langle \left( \hat{\phi}(\tau, m) \right)^2 \right \rangle &=& 
    (2 + 4 (m-1)) \sum_n \left \langle \left( \Delta A_n \hat{I_n}^z \tau / \hbar \right)^2 \right \rangle  \\
    &=& \frac{2 m - 1}{2} \sum_n \left \langle \left( \Delta A_n \tau / \hbar \right)^2 \right \rangle
\end{eqnarray}
since there are $2$ segments with a free evolution of $\tau$ and $m-1$ segments with a free evolution of $2 \tau$.
The net free evolution time is $t = 2 m \tau$.
For the special case of $T_2^*$, consider the Overhauser rotation with no refocusing pulses so that
\begin{equation}
\left \langle \left( \hat{\phi}(t, m=0) \right)^2 \right \rangle =
    \sum_n \left \langle \left(A_n \hat{I}_n^z t / \hbar \right)^2 \right \rangle
    = \frac{\sum_n A_n^2 t^2}{4 \hbar^2}.
\end{equation}

Using Eq.~\ref{eq_T2_sqrd} for $T_2^{(m)}$ with $m>1$ to denote CPMG with $m$ refocusing pulses and using $T_2^* = T_2^{m=0}$, we have
\begin{eqnarray}
\left(T_2^{(m)}\right)^2 &\leq& \frac{16 m^2 \hbar^2}{(2 m - 1) \sum_n \left \langle \left( \Delta A_n \right)^2 \right \rangle}, \\
\left( T_2^* \right)^2 &=& \frac{8 \hbar^2}{\sum_n A_n^2}, \\
\label{eq:zerr_proxy}
\left \langle \left(\frac{T_2^*}{T_2^{(m)}} \right)^2 \right \rangle
&\geq& \frac{2 m - 1}{2 m^2}
\left \langle \frac{\sum_n \left \langle \left( \Delta A_n \right)^2 \right \rangle}{\sum_n A_n^2} \right \rangle.
\end{eqnarray}

Ideally, the averaging in Eq.~\ref{eq:zerr_proxy} should be over a variety of dot-locations and/or devices.  In this way, the right side of Eq.~\ref{eq:zerr_proxy} will depend upon $\left \langle \left(\Delta \xi_i \right)^2 \right \rangle$ and $\left \langle \left(\frac{\partial \theta_v}{\partial \xi_i} \right)^2 \right \rangle$ for $i \in \{0, 1\}$ assuming independent distributions and averaging over pertinent $\theta_v$ function realizations.  At the sweet spot, the $\hat{Z} \otimes \hat{Z}$ error probability, to lowest order, depends upon 
$\left \langle \left(\Delta \xi_i \right)^4 \right \rangle$, $\left(\frac{\partial \theta_v}{\partial \xi_i} \right)^4$, $\left(\frac{\partial \theta_v}{\partial \xi_i} \right)^2 \frac{\partial^2 \theta_v}{\partial \xi_i^2}$, and $\left(\frac{\partial^2 \theta_v}{\partial \xi_i^2} \right)^2$; here, averaging is with respect to $\Delta \xi_i$ noise realizations but the $\theta_v$ function is fixed for a particular qubit.
We can relate $\left \langle \left(\Delta \xi_i \right)^4 \right \rangle$ to $\left \langle \left(\Delta \xi_i \right)^2 \right \rangle$ if we assume that $\Delta \xi_i$ are Gaussian distributed; then,
\begin{equation}
\label{eq:gauss_quartic}
\langle \left(\Delta \xi_i \right)^4 \rangle = \left \langle \left(\Delta \xi_i \right)^2 \right \rangle^2 \times (4-1)!! = 3 \langle \left(\Delta \xi_i \right)^2 \rangle^2.    
\end{equation}
Furthermore, we note that $\left \lvert \frac{\partial^2 \theta_v}{\partial \xi_i^2} \right \rvert \leq \left\langle \left(\frac{\partial \theta_v}{\partial \xi_i} \right)^2 \right\rangle$ should generally be true of smooth functions and uniform averaging~\footnote{It is easy to prove that 
$\left \lvert \frac{\partial^2 \theta_v}{\partial \xi_i^2} \right \rvert \leq \left\langle \left(\frac{\partial \theta_v}{\partial \xi_i} \right)^2 \right\rangle$ if $\theta_v$ is a sinusoidal function of $\xi_i$.  More generally, this property holds for any Fourier series, noting that the cross-terms of $\left(\frac{\partial \theta_v}{\partial x_0} \right)^2$ average to zero.}.
Finally, we may exploit our qubit selectivity freedom once more to choose qubits in which
$\left(\frac{\partial \theta_v}{\partial \xi_i} \right)^2 \leq \left \langle \left(\frac{\partial \theta_v}{\partial \xi_i} \right)^2 \right \rangle$ at its sweet spot; assuming $\frac{\partial \theta_v}{\partial \xi_i}$ is Gaussian distributed, about 68\% of candidates will satisfy this requirement.
Thus, with our assumptions, $\left(\frac{\partial \theta_v}{\partial \xi_i} \right)^4$, $\left(\frac{\partial \theta_v}{\partial \xi_i} \right)^2 \frac{\partial^2 \theta_v}{\partial \xi_i^2}$, and $\left(\frac{\partial^2 \theta_v}{\partial \xi_i^2} \right)^2$ are each bounded by a maximum of $\left\langle \left(\frac{\partial \theta_v}{\partial \xi_i} \right)^2 \right\rangle^2$.

We examine the worst-case performance at the sweet-spot ($c_0 = c_1 = 0$) by taking the limit of
$\left \lvert \frac{\partial^2 \theta_v}{\partial \xi_i^2} \right \rvert = \left\lvert\frac{\partial \theta_v}{\partial \xi_i} \right\rvert^2 \rightarrow \infty$ for $i \in \{0, 1\}$ and using the worst-case value of $\tan{\theta(z)}=1$ (given our $\tan{\theta(z)}<1$ qubit selectivity).  In this extreme limit and with our assumptions, from
Eqs.~(\ref{eq:zerr_via_deltaA}), (\ref{eq:A_perturbation}), (\ref{eq:perturb_sweetspot}), and (\ref{eq:gauss_quartic}) we derive
\begin{equation}
P_{Z \otimes Z}^{\rm err} \leq 3 \left( \frac{\pi}{2} \right)^2 \sum_{i=0}^1 \left( \left \langle \left(\frac{\partial \theta_v}{\partial \xi_i} \right)^2 \right \rangle \left \langle \left(\Delta \xi_i \right)^2 \right \rangle \right)^2.
\end{equation}
and from Eq.~\ref{eq:zerr_proxy},
\begin{eqnarray}
\nonumber
\left \langle \left(\frac{T_2^*}{T_2^{(m)}} \right)^2 \right \rangle &\geq& a \frac{2 m - 1}{2 m^2} \sum_{i=0}^1 \left \langle \left(\frac{\partial \theta_v}{\partial \xi_i} \right)^2 \right \rangle \left \langle \left(\Delta \xi_i \right)^2 \right \rangle \\
a &=& \left\langle \frac{\sum_n \tan^2{\theta_n} A_n^2}{\sum_n A_n^2} \right\rangle
\end{eqnarray}
We computed $a$ numerically for all combinations of silicon quantum dots with $r_0 \in \{10, 20\}~$nm and $z_0 \in \{5, 10\}~$nm for both 500~ppm and 1000~ppm $^{29}$Si and obtain $a = 0.34 \pm 0.01$ (bounding the estimate over the standard of error range in each case).

\begin{figure}
    \centering
    \includegraphics[width=\linewidth]{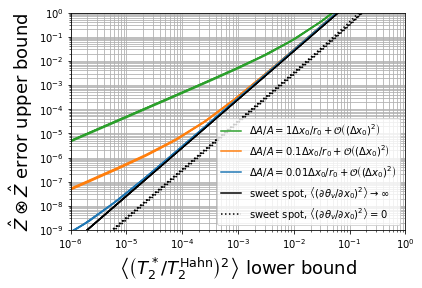}
    \caption{Upper bound of the $\hat{Z} \otimes \hat{Z}$ error probability versus the lower bound of $\left \langle \left(\frac{T_2^*}{T_2} \right)^2 \right \rangle$ under various conditions using the wavefunction model of Eq.~\ref{eq:wavefunction} and assumptions described in Sec.~\ref{sec:zz_err}.  
    The results are essentially identical for
    500~ppm and 1000~ppm $^{29}$Si and all four combinations of  $r_0 \in \{10~\textrm{nm}, 20~\textrm{nm}\}$ and $z_0 \in \{5~\textrm{nm}, 10~\textrm{nm}\}$.
    Results are shown both at and away from the sweet-spot.  Solid/dotted black curves are at a sweet spot in the pessimistic/optimistic limit with respect to valley phase contributions.
    Colored curves are away from the sweet spot with three different first-order contributions to $\Delta A / A$ [see Eq.~(\ref{eq:A_perturbation}) for context].
    \label{fig:zz_err_bound}}
\end{figure}

 We plot the pessimistic bounds of $\hat{Z} \otimes \hat{Z}$ versus $\left \langle \left(T_2^* / T_2^{\rm Hahn} \right)^2 \right \rangle$ at the sweet-spot for the extreme case of 
 $\left \lvert \frac{\partial^2 \theta_v}{\partial \xi_i^2} \right \rvert = 
 \left \lvert \frac{\partial \theta_v}{\partial \xi_i} \right\rvert^2 \rightarrow \infty$ and $\tan{\theta(z)}=1$ and the optimistic limit of
 $\frac{\partial^2 \theta_v}{\partial \xi_i^2} 
 = \left(\frac{\partial \theta_v}{\partial \xi_i} \right)^2 = 0$, as well as
 cases away from the sweet-spot, in Fig.~\ref{fig:zz_err_bound}.
 By using a ratio of $T_2$ versus $T_2^*$, the results are robust to isotopic enrichment and
 quantum dot size.
 As a point of reference $\left( T_2^* / T_2^{\rm Hahn} \right)^2 \approx 10^{-5}$ has been measured in SiGe devices~\cite{Eng2015,Kerckhoff2021}, but not one with electrostatic gates providing the bidirectional control that we require for the sweet-spot performance.
If we assume an accuracy of control that matches the reliability demonstrated by $T_2^{\rm Hahn}$ measurements, we can justify the lateral sweet-spot limit.  These results may easily be translated for longer CPMG pulse sequences according to the $m$ dependence in Eq.~\ref{eq:zerr_proxy}.  Using many-pulse CPMG may be valuable for making the bound tighter (via removing effects of nuclear flip-flops) and/or for probing longer timescales of the charge noise and temporal correlations of $\Delta A$.

\subsection{Overhauser field rotations}

During the relatively long duration of the e-n-CPhase operation, nuclei other than the Sn qubit may induce an unwanted rotation on the electron spin.  This extraneous nuclear spin bath imparts an effective magnetic field on the electron that is known as the Overhauser field.  In natural Si, nearly 5\% of the silicon atoms will have a nuclear spin.  These $^{29}$Si nuclear spins may be removed via enrichment which has been demonstrated in many Si qubit experiments~\cite{Eng2015,Kawakami2016,Yang2019,Kerckhoff2021}.  However, the cost of enrichment increases with the purity level and must be weighed against the benefits.  Furthermore, there may be other nuclear species present with nonzero spin depending upon the chemistry of the silicon well and the fabrication process.

In a modest magnetic field, nuclear spin baths are known to evolve slowly, largely through dipolar interactions among like nuclear species~\cite{Witzel2010, Witzel2012}.  Due to this fact, there are ways to mitigate this error.  It is possible to monitor the Overhuaser field and compensate for its effect (via ESR, spin-orbit~\cite{Prada2011spin,Jock2018,Tanttu2019} effects, or a micro-magnet~\cite{Kawakami2016}) as it slowly drifts~\cite{Bluhm2010}.  Furthermore, in principle, we can use a spin echo technique to filter out the low-frequency part of the Overhause field noise by flipping the extraneous spins relative to the qubit via NMR.  This could be effected by flipping just the bath spins or by flipping both the Sn and electron spins (but not the bath spins), and it may be performed while the qubits are interacting or in between two halves of the e-n-CPhase operation.

The electron Z-flip error induced by an Overhauser field is a simple function of $T_2^*$ (or effective $T_2^*$ if a mitigation strategy is employed):
\begin{eqnarray}
\nonumber
    P_{\rm e-Z-flip} &=& \left\langle \sin^2{(\hat{\phi}/2)} \right\rangle = \frac{1}{2} \left(1 - \exp{\left(-\left\langle \hat{\phi}^2 \right\rangle / 2\right)} \right) \\
\label{eq:overhauser_err}
    &=& \frac{1}{2} \left(1 - \exp{\left(-\left(T / T_2^* \right)^2\right)} \right)
\end{eqnarray}
where $T$ is the gate time.  This follows from Eq.~\ref{eq_T2_sqrd} via $\left\langle \hat{\phi}^2 \right\rangle = 2 \left(t / T_2^*\right)^2$ and the assumption (as before) that the outcomes of $\hat{\phi}$ are Gaussian-distributed.

\begin{figure}
    \centering
    \includegraphics[width=\linewidth]{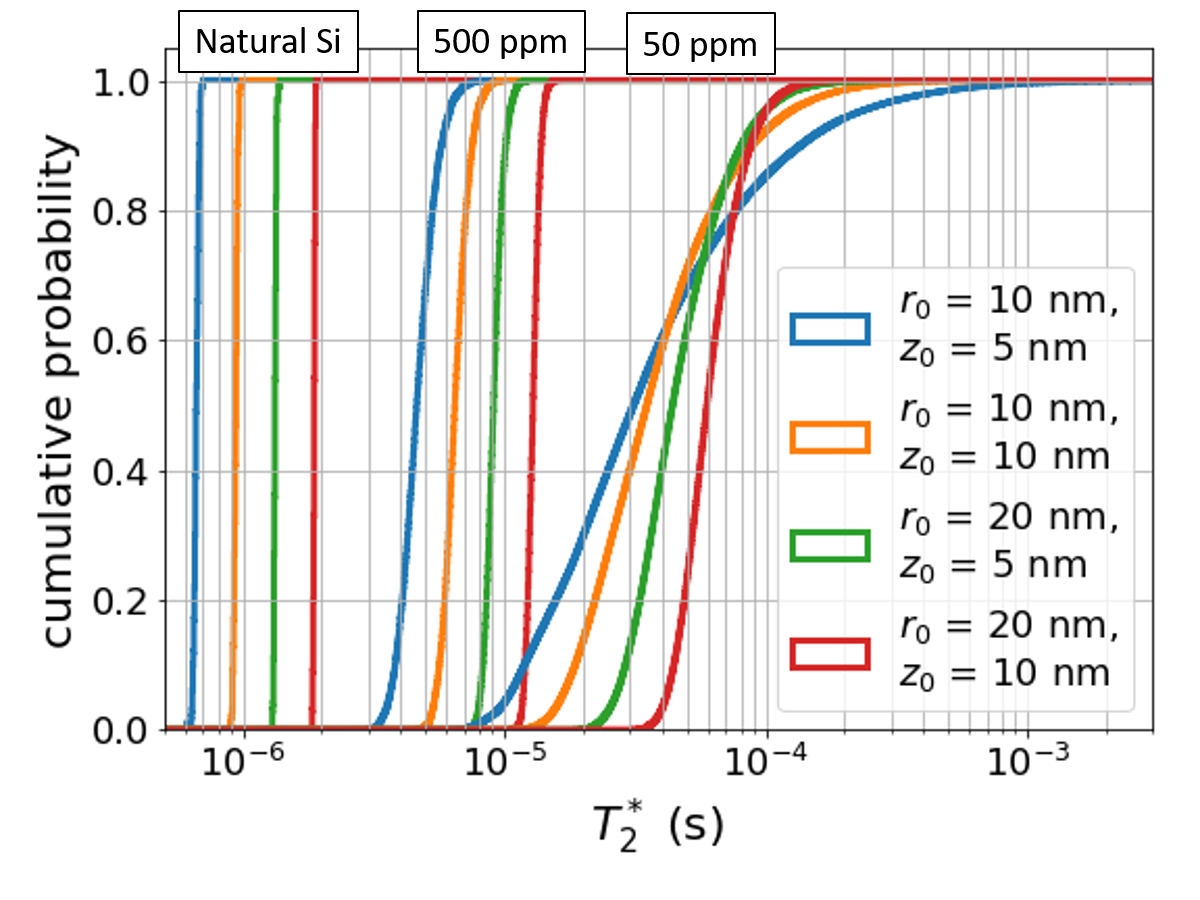}
    \caption{Cumulative probability distributions of $T_2^*$ for various quantum dot shapes and enrichment levels of $^{29}$Si (natural, 500 ppm and 50 ppm) using the simple model of Eq.~(\ref{eq:wavefunction}).  Other nuclear spins are not being considered here but may be important depending upon the material.
    \label{fig:T2star_cumprob}}
\end{figure}

To convey a sense of the potential magnitude of this error, Fig.~\ref{fig:T2star_cumprob} shows theoretical cumulative probability distributions of $T_2^*$ for various quantum dot shapes and levels of $^{29}$Si enrichment (with no other nuclear isotope considered) and Table~\ref{tab:probs_vs_T2star} presents error probabilities corresponding to a few $T_2^*$ values and gate times computed directly from Eq.~\ref{eq:overhauser_err}.  Without significant enrichment and/or a particularly strong hyperfine interaction with the qubit, this error may be substantial and concerning; however, it should be possible to reduce the effective $T_2^*$ considerably by employing either of the two mitigation strategies we have suggested above (spin-bath refocusing or drift compensation via tracking the Overhauser rotation).

\begin{table}[]
    \centering
    \begin{tabular}{c|ccc}
    \hline\hline
        $A h$ & $100$~kHz & $200$~kHz & $400$~kHz  \\
        $T$ & $5$~$\mu$s & $2.5$~$\mu$s & $1.25$~$\mu$s \\
        \hline
        \hline
        $T_2^*$ & & & \\
\hline 
       1~$\mu$s   &   0.5   &  0.5 &  0.4  \\
       10~$\mu$s   &   0.1   & 0.03 & 7.8$\times10^{-3}$  \\
       100~$\mu$s   &   1.3$\times10^{-3}$   &  3.1$\times10^{-4}$ & 7.8$\times10^{-5}$  \\
       \hline \hline
    \end{tabular}
    \caption{Electron Z error probabilities induced by Overhauser rotations for various $T_2^*$ values and e-n-CPhase gate times (with corresponding hyperfine strengths in Hz, $A h$ where $h$ is Planck's constant) computed directly from Eq.~\ref{eq:overhauser_err}.}
    \label{tab:probs_vs_T2star}
\end{table}

\section{Discussion and conclusion}
\label{sec:discuss}

Based on our calculations, there are compelling reasons to pursue the development of a technology for quantum information processing based upon Sn nuclear spins as qubits entangled by electrons that are shuttled through arrays of quantum dots.  We propose a simple and robust e-n-CPhase gate operation between Sn and electron qubits by shuttling an electron onto the Sn and waiting for a controlled-$\pi$ rotation in phase.  When combined with NMR and ESR qubit rotations, this is universal for quantum computing.  Using DFT to compute the bunching factor, $\eta$, for Sn in Si, we estimate that the Sn hyperfine interaction will be 10 times larger than $^{29}$Si.  This larger interaction translates to shorter e-n-CPhase gate times (a few $\mu$s) and a reduction of Overhauser error effects from extraneous nuclear spins.

We have analyzed the important error channels for this two-qubit operation: diabatic flip-flops, correlated $\hat{Z} \otimes \hat{Z}$, and Overhauser field rotations.  The first two can, in principle, have very low error probabilities, below $10^{-6}$, with a modest B-field ($> 15$~mT) and sufficiently precise control of the quantum dot onto a sweet spot that maximizes the hyperfine interaction.

We employ a novel technique to infer an upper bound on the correlated $\hat{Z} \otimes \hat{Z}$ over a $T_2$ timescale based upon averaging $\left(T_2^* / T_2\right)^2$ over a variety of device locations under the reasonable assumption that the valley phase is predominantly a function of the controllable quantum dot location. 
 Using values of $T_2$ and $T_2^*$ reported in the literature~\cite{Eng2015}, the error bound is about $5 \times 10^{-8}$ assuming that the control precision is as good as the reproducibility demonstrated by $T_2$.  However, missing the sweet spot target due to limitations of the control (e.g., precision or range) can increase this error probability significantly.  Beyond the $T_2$ timescale, regular characterization and compensation of drifting charge noise may be necessary to remain at the sweet spot and minimize this error channel probability.

The error coming from Overhauser field rotations is easily determined by the $T_2^*$ time relative to the gate time (which is inversely proportional to the hyperfine interaction strength with the qubit).  This error is expected to be significant without substantial enrichment or employing a mitigation strategy.  We suggest two mitigation strategies: tracking the Overhauser field rotation and compensating for its drift; spin refocusing by flipping the bath spins relative to the qubit spins.

An experimental realization of this system will involve a quantum dot array in silicon with an integrated platform for NMR and ESR, similar to what has already demonstrated (see, e.g. Ref. \onlinecite{Hensen2019}). Successful incorporation of Sn atoms into silicon quantum dots with minimal quantum dot degradation will need to be demonstrated. Tin is soluble in silicon at levels up to $x\sim0.16\%$~\cite{Trumbore1960}, which should provide an adequate number of Sn atoms in any given quantum dot for these qubits to be studied. 

We find the possible combination of NMR-based high-fidelity single qubit operation and high-fidelity nuclear entanglement operation in the Sn in silicon system compelling. 
Experiments will be crucial in testing the ideas presented here and whether the remarkable prospects of quantum-dot-coupled tin qubits in silicon are realized.

\section{Acknowledgements}
We acknowledge N. Tobias (Toby) Jacobson and Ryan Jock for valuable feedback, questions, and insights, and we thank Quinn Campbell for reviewing our DFT methodology.
We are also grateful for an allocation on Sandia National Laboratories  High Performance Computing resources and for exceptional technical support from Phillip Regier. Sandia National Laboratories is a multimission laboratory managed and operated by National Technology \& Engineering Solutions of Sandia, LLC, a wholly owned subsidiary of Honeywell International Inc., for the U.S. Department of Energy's National Nuclear Security Administration under contract DE-NA0003525. This paper describes objective technical results and analysis. 
Any subjective views or opinions that might be expressed in the paper do not necessarily represent the views of the U.S. Department of Energy or the United States Government.

\section{Appendix: Relativistic contact density corrections}

Computed contact densities were used in Sect. \ref{sec:DFT} to generate 
bunching factors and HFIs for the Group-IV silicon-substitutional defects
$^{29}$Si, $^{73}$Ge, and $^{119}$Sn, and the predicted values for 
$\eta_{\rm Si}$ and $\eta_{\rm Ge}$ agreed very well with measurements.
Meanwhile, other Group IV nuclides, including $^{13}$C and $^{207}$Pb, 
were omitted from consideration due to the unavailability of measured 
reference values. In this appendix, we compute $\eta$ values for the 
full series of Group IV substitutional defects C--Pb and determine
whether known $Z$-scaling manifests. With Pb being a close neighbor 
of Au, the local maximum of relativistic effects, 
we first address relativistic deficiencies in Fermi's contact density formula. 
After obtaining an improved set of $\eta$ values, we provide 
an updated prospectus for the various candidate spin-qubit defects.

Contact density scaling with atomic number has been studied extensively, 
with many functional forms proposed in the context of free atoms.
Within a non-relativistic formalism, popular examples include 
an analytic quantum defect theory model developed by Blinder \cite{Blinder1979},
a numerical Hartree-Fock-based model of Koga et al.\cite{Koga1995},
and the analytic asymptotic analysis of Heilmann and Lieb \cite{Heilmann1995},
just to name a few.
Here our interest extends beyond the non-relativistic regime, and as such 
the considerations of Otten \cite{Otten1989} are more appropriate. 
In the context of the HFI, he applied several relativistic corrections 
to the Fermi contact interaction and concluded that the HFI scales 
with $Z$ and the atomic mass $A$ as $Z^2$/$A^{1/3}$. In another context, 
namely for contact-density derived atomic and molecular field shifts, 
one of the authors found this scaling formula to be applicable through at least $Z$=70
\cite{Lutz2016}.

It is straightforward to correct Eq.\ \ref{eq:FCI} to leading order for relativity.
Following publication of Fermi's celebrated non-relativistic derivation 
of the contact interaction \cite{Fermi1930}, a correction accounting for
relativity was provided by Breit\cite{Breit1930}. Breit's formula was 
later generalized by Inokuti and Usui \cite{IU1957} to hydrogen-like 
orbitals with arbitrary 
principle quantum number $n$ and presented as a series expansion:
\begin{equation}
    B(n,Z) = 1 + \frac{n^2 + 9n - 11}{6n^2} (\alpha Z)^2 + O(\alpha^4).
    \label{eq:IU}
\end{equation}
This result was discovered independently by Pyykk{\"o} \cite{Pyykko1971} 
who later justified its use for multi-electronic atoms\cite{Pyykko1972,Pyykko1973}. 
Here we apply the correction given by Eq.\ \ref{eq:IU} to our own DFT-derived
contact densities, generated as described in the main text
\footnote{The fact that our contact densities were generated using relativistic DFT is a separate issue; Eq.\ \ref{eq:FCI}, which was derived within a non-relativistic framework, becomes relativistic to first-order by application of the correction given by Eq.\ \ref{eq:IU}}.

\begin{figure}[b!]
    \centering
    \includegraphics[width=\linewidth]{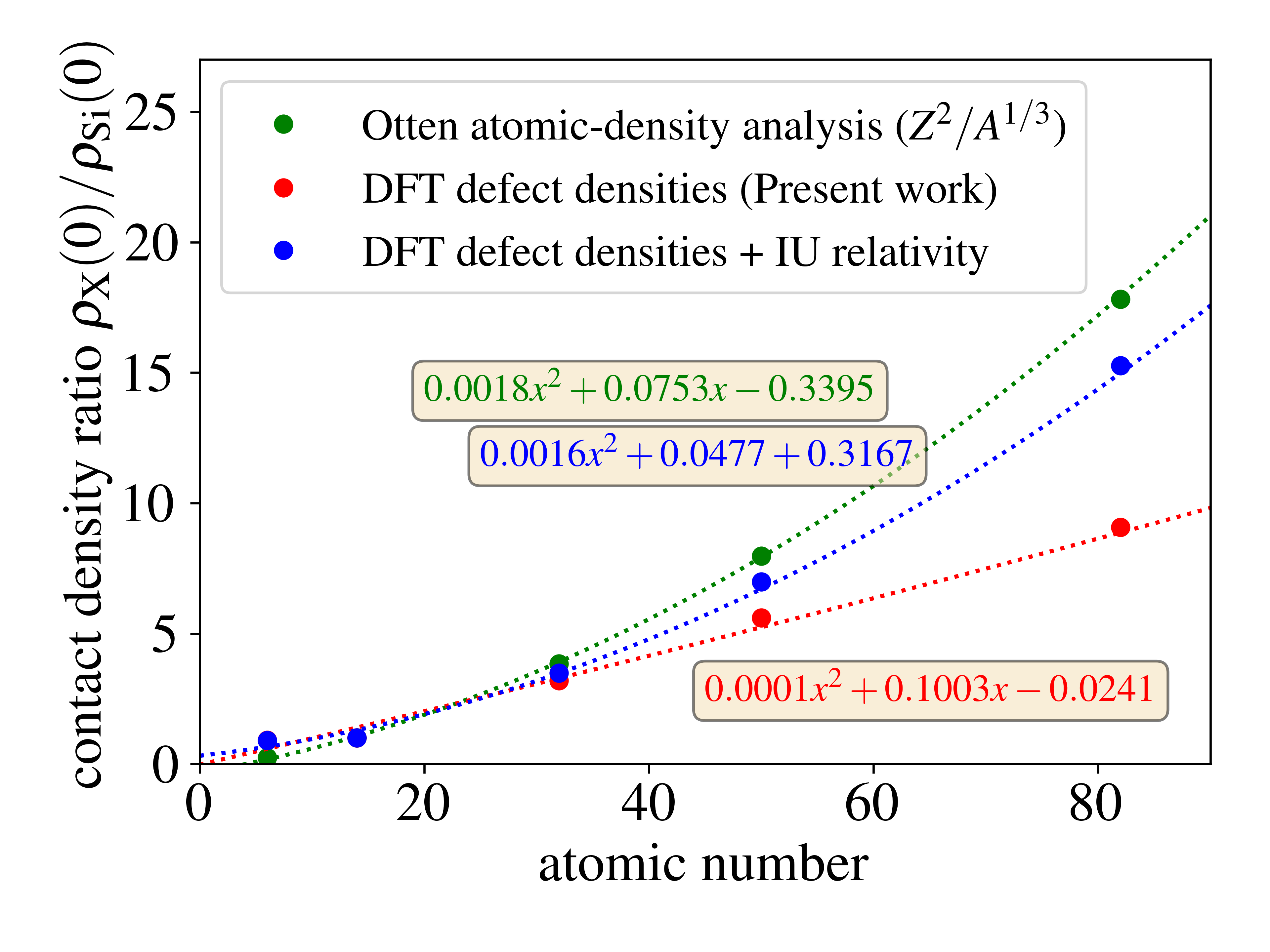}
    \caption{Contact density ratios for Group IV Si-substitutional defects
    taken with respect to the value for intrinsic $^{29}$Si. 
    DFT values presented in the main text are shown with and without
    inclusion of the relativistic correction given in Eq.\ \ref{eq:IU}.
    For reference, we also show values generated using the relativistic 
    scaling law of Otten \cite{Otten1989}.}
    \label{fig:appendix}
\end{figure}

Figure \ref{fig:appendix} collects contact density ratios, taken with respect to Si,
for Group-IV Si-substitutional defects C to Pb. 
We show our {\it ab initio} data with and without the relativistic correction of Inokuti and Usui (IU).
For comparison, we also present contact density ratios from the analytic model for 
free atoms by Otten for the same Group-IV series.
Fits to quadratic functions returned $R^2$ values
of 0.992, 0.999, and 0.9999 for the DFT, IU-corrected DFT, and Otten contact densities, 
respectively. 
Excellent agreement is observed between our IU-corrected DFT values and
the relativistic Otten scaling. 
Perfect agreement is not expected because these are modeling different
scenarios (defects in Si versus free atoms), but it is reassuring to observe similar
behavior where the atoms at the contact site are the same.
After adjusting the Sn FCI values appropriately, 
the new absolute HFI is 2400. Again when taken with respect to intrinsic Si, 
the HFI enhancement for Si:Sn increases to $13.5*\eta_{\rm Si}$ from its 
`non-relativistic' value of $10*\eta_{\rm Si}$ given in the main text.

Finally, whether or not it is possible to fabricate Si:Pb in practice, we infer 
its prospects here in light of computations shown in Figure \ref{fig:appendix}. 
The IU-corrected DFT value of $\eta_{\rm Pb}:\eta_{\rm Si}=15.3$ translates to 
an absolute HFI of 2920, or a $^{207}$Pb:$^{29}$Si HFI ratio of only 16.4.
Comparing this to the $^{119}$Sn:$^{29}$Si HFI ratio of 13.5, 
the Pb enhancement is only $\sim$~20\% larger. Taking into account the negligible 
solubility of Pb in Si, it is unclear whether chasing this additional HFI 
enhancement will be worthwhile.

\end{document}